\documentclass[%
 aip,
 amsmath,amssymb,
 reprint,%
]{revtex4-1}
\usepackage{graphicx}
\usepackage{dcolumn}
\usepackage{bm}

\usepackage[utf8]{inputenc}
\usepackage[T1]{fontenc}
\usepackage{mathptmx}
\usepackage{etoolbox}
\usepackage{siunitx}
\usepackage{ulem}
\usepackage{prettyref}						
\newrefformat{fig}{Figure~\ref{#1}}
\usepackage[dvipsnames]{xcolor}
\makeatletter
\def\@email#1#2{%
 \endgroup
 \patchcmd{\titleblock@produce}
  {\frontmatter@RRAPformat}
  {\frontmatter@RRAPformat{\produce@RRAP{*#1\href{mailto:#2}{#2}}}\frontmatter@RRAPformat}
  {}{}
}%
\makeatother
\begin{document}

\preprint{AIP/123-QED}

\title{Monolithic optoelectronic circuit design for on-chip terahertz applications}
\author{Kateryna Kusyak}
\affiliation{Max Planck Institute for the Structure and Dynamics of Matter, Hamburg, Germany}
 \affiliation{Department of Physics, Columbia University, New York, NY, USA}
 \affiliation{Center for Free-Electron Laser Science, Hamburg, Germany}
\author{Benedikt Schulte}%
\affiliation{Max Planck Institute for the Structure and Dynamics of Matter, Hamburg, Germany}
 \affiliation{Department of Physics, Columbia University, New York, NY, USA}
 \affiliation{Center for Free-Electron Laser Science, Hamburg, Germany}
\author{Toru Matsuyama}
\affiliation{Max Planck Institute for the Structure and Dynamics of Matter, Hamburg, Germany}
\affiliation{Center for Free-Electron Laser Science, Hamburg, Germany}
\author{Gunda Kipp}
 \affiliation{Max Planck Institute for the Structure and Dynamics of Matter, Hamburg, Germany}
 \affiliation{Center for Free-Electron Laser Science, Hamburg, Germany}
\author{Hope M. Bretscher}
 \affiliation{Max Planck Institute for the Structure and Dynamics of Matter, Hamburg, Germany}
  \affiliation{Department of Physics, Columbia University, New York, NY, USA}
  \affiliation{Center for Free-Electron Laser Science, Hamburg, Germany}

\author{Matthew W. Day}
\affiliation{Max Planck Institute for the Structure and Dynamics of Matter, Hamburg, Germany}
 \affiliation{Department of Physics, Columbia University, New York, NY, USA}
 \affiliation{Center for Free-Electron Laser Science, Hamburg, Germany}
 \author{Guido Meier}
\affiliation{Max Planck Institute for the Structure and Dynamics of Matter, Hamburg, Germany}
\affiliation{Center for Free-Electron Laser Science, Hamburg, Germany}
 \author{Alexander M. Potts}
\affiliation{Max Planck Institute for the Structure and Dynamics of Matter, Hamburg, Germany}
 \affiliation{Department of Physics, Columbia University, New York, NY, USA}
 \affiliation{Center for Free-Electron Laser Science, Hamburg, Germany}
\author{James W. McIver}
\affiliation{Max Planck Institute for the Structure and Dynamics of Matter, Hamburg, Germany}
 \affiliation{Department of Physics, Columbia University, New York, NY, USA}
 \affiliation{Center for Free-Electron Laser Science, Hamburg, Germany} \email{jm5382@columbia.edu.}

\date{\today}

\begin{abstract}

We demonstrate a monolithic coplanar stripline platform for on-chip terahertz (THz) generation, transmission, and detection, addressing key challenges of mode purity, bandwidth, and referencing. Capacitive coupling of the photoconductive generator switch enforces pure odd-mode propagation, increases THz field strength, and extends the operational frequency range, achieving 0.05–1.4 THz. Our architecture enables fully monolithic fabrication with amorphous silicon switches, \textit{in situ} field referencing, and galvanic isolation between generation and detection. Finite-element simulations and experiments confirm that suppressing parasitic modes improves signal integrity, providing a robust platform for high-fidelity THz spectroscopy, ultrafast electronics, and nanoscale quantum materials research.

\end{abstract}

\maketitle
\section{\label{sec:level1}Introduction}

\begin{figure}
\includegraphics{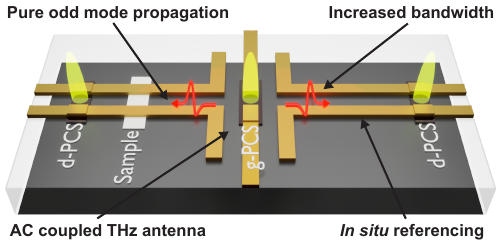}
\caption{Schematic of the THz optoelectronic circuitry. In the center, a near-field THz generator in the form of a photoconductive switch (g-PCS) is connected with two electrodes. The generation is capacitively coupled to the antennas, linked to coplanar striplines.  Detection switches (d-PCS) are placed on each side, \SI{2}{mm} from the g-PCS. The advantages of this design are the propagation of only the odd mode, an increase in bandwidth, \textit{in situ} referencing for on-chip THz spectroscopy and galvanic isolation of generation and detection.}
\label{fig:1}
\end{figure}

The rapid advance of THz technology has significantly broadened its application in spectroscopy, communications, and materials science. Positioned between radio and optical frequencies, THz technology offers practical advantages over microwave electronics, including higher data rates\cite{Ishigaki2012} and lower latency for enhanced communication,\cite{Rappaport2019} while enabling high-resolution imaging and spectroscopy of chemical,\cite{Mittelmann1998, Weide2000, Federici1005} biological,\cite{Nagel2006} and medical systems.\cite{Siegel2004, Woodward2003} In condensed matter physics and materials science, the THz regime is crucial for studying many-body collective modes which typically exhibit characteristic absorption profiles in this frequency range.\cite{Basov2011, Basov2017, Torre2021} The comparatively low energy scale of THz radiation means it can function as a non-destructive material probe. 
Advances in on-chip THz optoelectronics have enabled near-field THz spectroscopy of nano- and micrometer-sized systems. The diffraction limit at 1 THz is $\sim$\SI{150}{\micro m}, which is more than an order of magnitude larger than the typical size of biological systems or two-dimensional van der Waals materials, usually ranging between hundreds of nanometers and a few micrometers.\cite{Nagel2006, Basov2011} The far-field diffraction limit can be circumvented by confining THz radiation using circuitry that unites transmission lines with femtosecond laser-triggered photoconductive switches (PCSs).\cite{Auston1975} The sub-picosecond carrier lifetimes and high on-off contrast ratio of semiconductor PCSs enable them to serve as both generators and detectors of THz radiation. The PCSs transmute an ultrafast optical pulse into a sub-picosecond current pulse, which is coupled into a transmission line that can be interfaced with downstream electrical components, samples, or other THz active systems. In order to propagate in-plane THz fields, coplanar striplines (CPS) are commonly used.\cite{Gallagher2019,Zhao2023, Smith2019, Potts2023, Yoshioka2022, Yoshioka2024,Seo2024} 

Conventional CPS-based THz circuit designs, however, face significant limitations. Traditional layouts are known to launch multiple propagating eigenmodes, which complicates spectroscopic or electronic applications and can reduce the device bandwidth. \cite{Kipp2024,Karnetzky2018} This multi-mode behavior arises because the PCS is typically side-coupled to the signal line only of the coplanar striplines, leading to mixed mode excitation.\cite{, Wu2015} Additionally, effective on-chip THz spectroscopy requires independent control over the quiescent biases of the generator and detector switches — a task that is difficult in continuous CPS geometries without adding extra biasing lines, which in turn can parasitically disturb the propagating THz modes.
Finally, many existing circuit designs\cite{wood2013, Gallagher2019,Zhao2023,Smith2017, Smith2019, Hunter2015, Potts2023, Yoshioka2022, Yoshioka2024} use III-V compound semiconductors like LT-GaAs as PCSs. Such materials are grown on substrates which are lossy at THz frequencies. Thus, epitaxial lift-off and transfer of the III-V material onto substrates suitable for THz frequencies become mandatory. Those procedures require hazardous chemicals and dry transfer, introducing complexity and therefore reducing nanofabrication precision and throughput.

In this work, we present an improved THz optoelectronic circuit design that addresses the key challenges of pure mode propagation, limited bandwidth, \textit{in situ} referencing, and galvanic isolation. We achieved this by developing a CPS architecture with a capacitive THz pulse generator, which couples silicon-based PCSs directly to the transmission lines. This approach enforces pure odd-mode propagation, enhances bandwidth through optimized coupling, and enables monolithic fabrication with integrated absolute referencing. Compared to conventional side-coupled designs, our circuitry extends the operational frequency range, achieving 0.05–1.4 THz bandwidth. By overcoming traditional limitations, our design provides a versatile and accessible platform for high-throughput, on-chip THz spectroscopy of nanoscale systems.

\section{\label{sec:results}Results}

\subsection{\label{sec:design} Design}

\prettyref{fig:1} shows a schematic of our AC-coupled circuit design for signal transmission. We use amorphous silicon ($\alpha$-Si) for PCSs, which is known to be a reliable THz emitter.\cite{Zhong2008, Karnetzky2018, McIver2020, Island2020, Wang2023, Adelinia2025, Kipp2024} The PCSs (black patches in \prettyref{fig:1}) are  deposited onto a \SI{2}{mm} thick sapphire substrate (Al$_2$O$_3$, c-cut) via electron beam evaporation. 
The generator PCS (g-PCS) was contacted by two electrodes, across which a DC bias is applied.  A \SI{310}{\femto s} laser pulse, centered at \SI{515}{\nano m}, illuminated the g-PCS with a \SI{196}{kHz} repetition rate modulated at \SI{98}{kHz} by an acousto-optic modulator. The pulses are capacitively (AC) coupled to the CPS. The details of the coupling are discussed below, in \prettyref{sec:coupling}. This design acts as a DC block between g-PCS and detector PCSs (d-PCSs), which allows all PCSs to be between the metal traces. 
Symmetric transmission lines extend for $\sim$\SI{5}{mm} away from the g-PCS on both sides. d-PCSs are placed \SI{2}{mm} away from the g-PCS along the CPS. The round trip reflections from the bond pads are separated by \SI{20}{ps} and the lower frequency cutoff of the circuit is, therefore, \SI{50}{\giga \hertz}. This layout enables placement of a planar sample between the two metal traces, \SI{1}{mm} from the g-PCS, while leaving the opposite side bare to provide a reference signal propagating through an unaltered CPS. 

\subsection{\label{sec:even&odd}Mode propagation}
\begin{figure*}
    \includegraphics{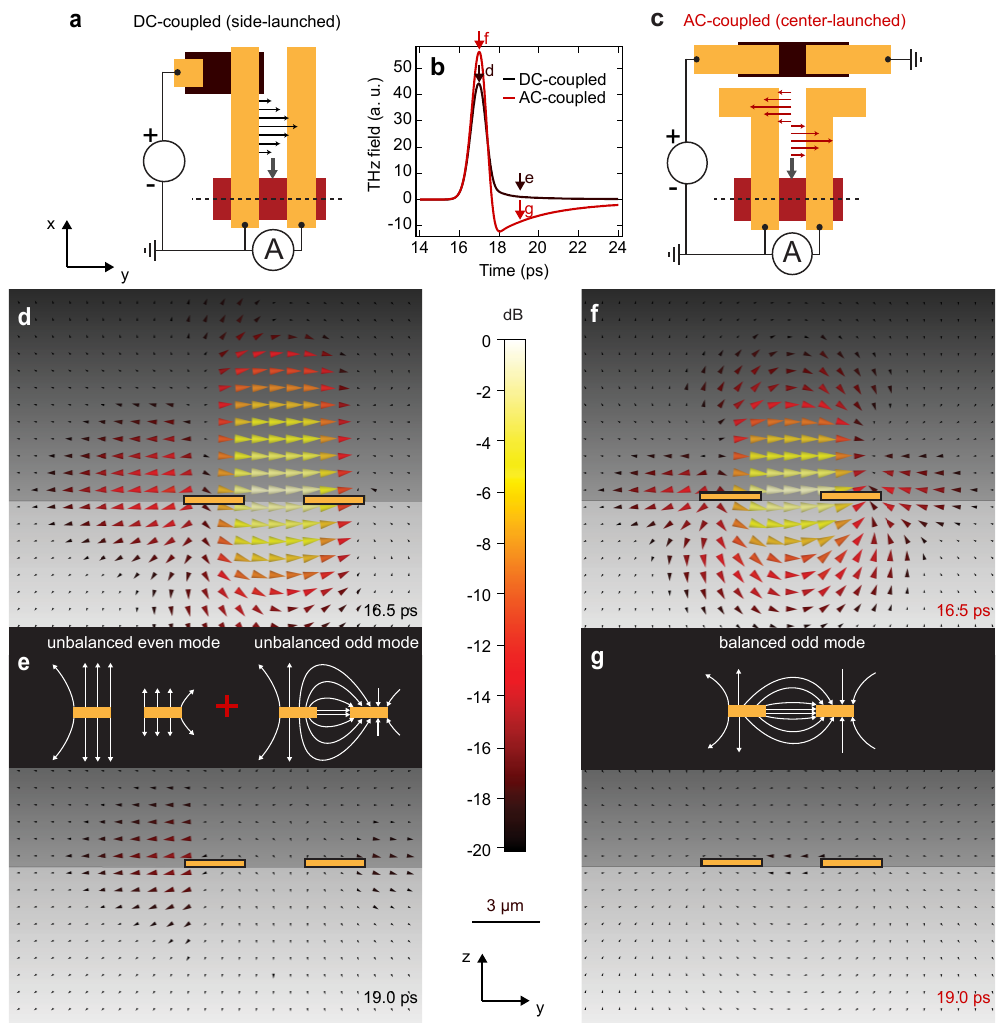}
    \caption{(a) Schematic of the conventional design of on-chip THz spectroscopy with DC coupled side-launched generation. (b) Numerical time-domain simulation of DC (black) and AC (red) coupled designs. (c) Schematic of our design with AC-coupled center-launched generation. (d)-(g) Cross section of electric field distributions at the d-PCS after \SI{2}{mm} pulse propagation, normalized to dB. (d) and (e) show the fields of the DC-coupled design, and (f) and (g) of the AC-coupled design at \SI{16.5}{ps} and \SI{19}{ps}, respectively. Insets of (e) and (g) show schematics of propagating modes on CPS.}
    \label{fig:2}
\end{figure*}
The CPS geometry supports several quasi-TEM modes, including the even mode, the odd mode, and linear combinations of the two. In the even mode, the charge distribution is symmetric across the metallic lines, and the resulting electric field points primarily perpendicular to the plane of the CPS. In contrast, the odd mode features an antisymmetric charge distribution, with electric field lines concentrated between the metal traces, pointing in-plane. The mode profiles are further complicated if the condition of balanced propagation — equal and opposite charge distributions on the two conductors — is relaxed. In this case, unbalanced modes can arise, characterized by asymmetric charge distributions between the metallic traces. These unbalanced modes distort the electric field profile, leading to leakage into surrounding structures, electromagnetic interference, and crosstalk between adjacent circuit elements. Suppressing unwanted modes is critical for preserving signal integrity, minimizing dispersion, and ensuring reliable performance across a wide range of THz optoelectronic devices.\cite{simonovich2022}

We systematically explored the generation and propagation of even and odd modes in our AC-coupled CPS circuit, as well as in the more commonly used side DC-coupled geometry. To attain the THz field distributions of each design, we performed numerical finite integration technique (FIT) simulations using CST Studio Suite.\cite{CSTmanual}
\prettyref{fig:2} summarizes our computational findings. In the side-launched, DC-coupled design (\prettyref{fig:2}(a)), the g-PCS is ohmically contacted to one of the CPS metal traces. A THz pulse is generated by photoexciting the biased g-PCS (black region). In contrast, in the center-launched, AC-coupled design (\prettyref{fig:2}(c)), the g-PCS is galvanically isolated from the CPS and biased differentially across two electrodes. In both configurations, as the THz transient propagates past the d-PCS, the instantaneous THz field acts as a transient voltage bias across the d-PCS. When the d-PCS is illuminated by a gate pulse, this bias drives a photocurrent whose amplitude is proportional to the local THz field, and can be measured using a transimpedance amplifier and standard lock-in techniques, documented elsewhere.\cite{McIver2020}

\prettyref{fig:2}(b) shows the simulated electric field in the plane of the CPS for both the DC- and AC-coupled designs. For both designs, a current pulse
\begin{equation}
    I(t) = I_0 ~ \exp\left(-4 \ln(2)\left(\frac{t-t_0}{\text{FWHM}}\right)^2\right)
\end{equation}
was applied to the surface of the g-PCS. The full width half maximum (FWHM) of the Gaussian pulse shape was set to \SI{0.82}{ps} to match the experimental pulse width and the center of the pulse $t_0=\SI{1.764}{ps}$ was picked arbitrarily to resemble the data. The Gaussian pulse shape was chosen as an approximation of the photo-induced current in the g-PCS and yields good agreement with the experimental data. The simulated Gaussian pulse is the simplest model pulse which induces the photocurrent in the g-PCS and whose Fourier Transform possess a DC component.
The time-dependent THz field was evaluated after \SI{2}{\milli m} propagation, corresponding to g-PCS and d-PCS distance. The black dotted lines in panels (a) and (c) mark the locations where the electric field profiles shown in panels (d)–(g) were extracted. The widths and gap between the metal traces of the CPS were all set to 3~$\mu$m and the height of the traces to 275~nm. From the numerical simulation we obtained the effective dielectric constant between the chosen sapphire substrate and vacuum of $\varepsilon_{\text{eff}}=5.1$.
 The THz propagation in the CPS, thus, was estimated to be $\sim 0.44~c= 132~\mu$m/ps, which is in good agreement with the measured velocity.\cite{Kusyak2022}
This velocity corresponds to $\sim$\SI{16}{ps} propagation after the generation process. \prettyref{fig:2}(d) – (g) show electric field distributions at the cross section of the CPS after a propagation of \SI{2}{mm}. The arrows in panel (b) indicate the points in time in which the field distributions are represented. The evolution of the fields for both designs at additional points in time can be found in the supplementary material. Panels (d) and (e) are field distributions for DC-coupled circuits, (f) and (g) for AC-coupled circuits. All fields are normalized to the range of \SIrange{0}{-20}{dB} for ease of comparison.

In the DC-coupled, side-launched design, a larger fraction of the electric field remains confined between the CPS traces, but a significant portion radiates out of the structure due to the asymmetrical placement of the g-PCS. The field shown in \prettyref{fig:2}(e) corresponds to propagation of a mixture of both unbalanced even and odd modes, indicated by the isolated schematic of CPS propagation modes in the inset. In \prettyref{fig:2}(e), \SI{19}{ps} after excitation, the consequences of unbalanced even and unbalanced odd mode generation are more severely manifested, as the majority of the electric field is unbalanced and propagating out of the metallic traces instead of between them. Meanwhile, the field distribution in the AC-coupled design (\prettyref{fig:2}(f)) shows an electric field resembling that of a charge dipole, indicating pure balanced odd mode propagation. For comparison, the inset of (g) shows the corresponding schematic. Multi-mode generation is limited as a result of the g-PCS and d-PCS positioned between the metal traces and thus the transient current generated between the traces predominantly excites the odd mode.
Moreover, at later points in time, the field is not propagating outside of the CPS, but is only present between the metal traces, in stark contrast to panel (e) (as can be seen in the field distribution videos in the supplementary material). Another aspect of multiple propagating modes to consider is that different characteristic impedances and phase velocities lead to dispersion and, therefore, temporal broadening. Preventing multi-mode propagation, thus, is also advantageous in terms of bandwidth and ease of circuit design.

Note, while optimized geometries for CPS are shown here, field distributions of both DC and AC-coupled design with \SI{8}{\micro m} conductor widths and separation gaps can be found in the supplementary material. The signal of the \SI{8}{\micro m} design is attenuated compared to that of the \SI{3}{\micro m} design, due to the skin-effect,\cite{Wheeler1942} and the mode mixing of the corresponding DC-coupled circuit is more pronounced. However, reducing the geometry below \SI{3}{\micro m} does not improve the single mode propagation significancy, but leads to a concentration of the field on the inner edges of the metal traces, increasing the attenuation. 

\subsection{\label{sec:coupling} Coupling}
\begin{figure}
    \includegraphics{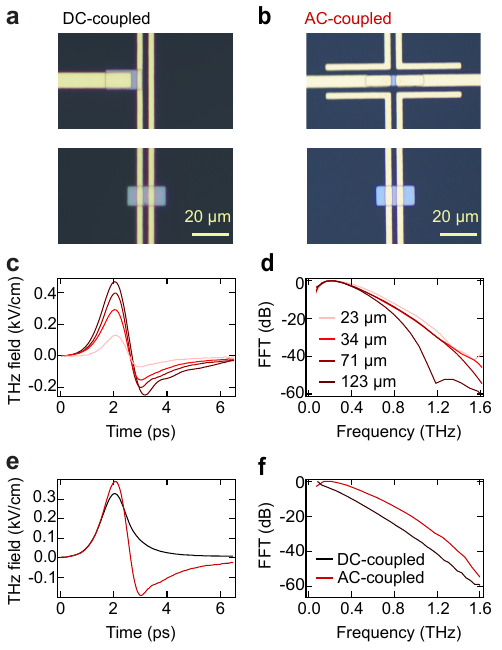}
    \caption{Micrographs of DC-coupled (a) and AC-coupled (b) circuits consisting of evaporated $\alpha$-Si and evaporated Ti:Au on sapphire substrates. (c) Measured THz field of circuits of different antenna lengths with the corresponding fast Fourier Transform in (d), normalized to dB.
    (e) THz field in the time-domain of the DC coupled (black) and AC-coupled (red) designs. (f) Fast Fourier Transform of the data shown in (e) normalized to dB. Data in panels (c)-(f) were taken at 7~K.}
    \label{fig:3}
\end{figure}

With our design we observed an increase in bandwidth and THz field amplitude. Generally, a challenge with transmission lines is their ohmic\cite{Grischowsky1990} and \v{C}erenkov\cite{Auston1984, Grishkowsky2000} losses, which can limit the operating bandwidth. Optimizing bandwidth and transmitted field are, thus, known goals in circuit design. We fabricated both DC-coupled and AC-coupled designs.
Micrographs of both circuits can be seen in \prettyref{fig:3}(a) and (b).
In \prettyref{fig:3}(b), the DC blocks on each side after the g-PCS, as well as the perpendicular extensions, which allow for the coupling, are shown. We refer to those extensions as antenna arms. The length of the shown antenna arms was \SI{34}{\micro m} each, resulting in a total length of \SI{71}{\micro m} (including the \SI{3}{\micro m} gap). 

From our measurements, we established \SI{34}{\micro m} per antenna arm as the best compromise between THz field amplitude and bandwidth. 
To determine coupling efficiency and bandwidth, we fabricated and tested different total antenna lengths, including gaps: \SI{23}{\micro m}, \SI{34}{\micro m}, \SI{71}{\micro m} and \SI{123}{\micro m}. Their measured electric field profiles can be seen in \prettyref{fig:3}(c) in the time domain and in \prettyref{fig:3}(d) in the frequency domain. The frequency domain data is scaled so the maximum is \SI{0}{dB}. 
Two main effects of the antenna length can be observed. First, longer antennas lead to an increase in the amplitude of the propagating THz field. This can be explained by the larger coupling area between g-PCS electrodes and the AC-coupled CPS. The coupling strength, however, is non-linear with antenna length and saturates, as can be seen in Figure~A-1(a) in the supplementary material. Second, the antenna length influences the spectral bandwidth (see \prettyref{fig:3}(d)).
With the antenna length of \SI{123}{\micro\meter}, the time-domain signal has an oscillatory behavior between 4-\SI{6}{ps} in \prettyref{fig:3}(c). Assuming a propagation speed of 0.44$\cdot c$, the oscillation corresponds to the resonance at \SI{1.17}{THz} seen in \prettyref{fig:3}(d). The resonance matches the round trip time of light propagating to and reflecting from the ends of the antenna arms. We conclude that shorter antennas also induce reflections, but such resonances occur at frequencies beyond our resolvable bandwidth.

Using our preferred design, we compared the signal with the conventional DC-coupled design.
 \prettyref{fig:3}(e) displays the time domain signals, which show a half-cycle pulse for DC-coupled (black) and a full-cycle pulse for AC-coupled (red) designs.
Applying a fast Fourier Transform to the signals gives the spectra in \prettyref{fig:3}(f). The series capacitance of the AC-coupling can be modeled as a differentiator, indicated by the time-domain signals in \prettyref{fig:3}(e).
 The transmitted spectrum is shifted to higher frequencies as a result of the AC-coupling. The \SI{-3}{dB} point increases from \SI{160}{GHz} in the DC-coupled design to \SI{440}{GHz} in the AC-coupled design. The \SI{1}{THz} component of the signal is \SI{-39}{dB} and \SI{-23}{dB} in the DC-coupled and AC-coupled designs, respectively, compared to the frequency component with the strongest transmitted signal. This improves the signal to noise and extends the bandwidth of the AC-coupled design to higher frequencies.

Additionally, in the DC-coupled design a small electric field persists up to \SI{25}{ps} after the peak. In contrast, the capacitance of the AC-coupled design acts as a high-pass filter to this undesired artifact, tightening the wave packet in time. This enables shorter measurement windows without additional post-processing techniques. It also prevents low-frequency signals from reflecting and interfering with the extraction of desired signal, thus resolving lower bandwidth signals more reliably. 

A final advantage of the AC-coupled design, is the increase of transferred energy by a factor of $\sim$1.5. This is due to the better coupling efficiency with no energy wasted by transfer to the even mode, as seen by the larger field in \prettyref{fig:3}(e).

\subsection{\label{sec:}Spectroscopy}
We demonstrate that our monolithic THz circuit enables accurate, linear-response spectroscopy, supported by reliable device fabrication and integrated \textit{in situ} referencing.
Extracting the real and imaginary optical conductivity of a sample with on-chip THz spectroscopy is a clear use case for this improved circuit design.
The optical conductivity of a specific sample, however, can only be fully extracted if the contribution of the background can be isolated and removed. Previous experiments have used THz circuits to extract the conductivities of samples, using gate tunable van der Waals materials such as graphene. In those materials, \textit{in situ} carrier density control allowed a reference trace to be measured at the charge neutrality point, where the sample is transparent to THz.\cite{Gallagher2019, Zhao2023, Seo2024} This approach, however, does not account for coupling between the gate and van der Waals material\cite{Kipp2024} and precludes measuring samples without an \textit{in situ} tunable insulating state. These complexities are mitigated with an absolute reference.\cite{koch2023}

\begin{figure}
    \includegraphics{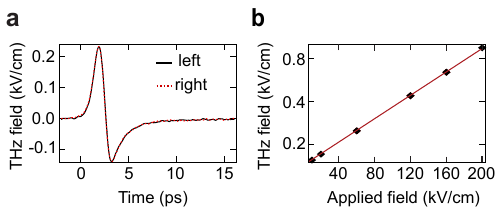}
    \caption{(a) Time-domain data comparing left (black) and right (red) detector switches on a single chip after calibration. (b) THz field dependence on generator bias. Data was taken at 7~K.}
    \label{fig:4}
\end{figure}

To accurately extract the propagated THz electric field amplitude, we calibrated the photoconductive switch response to laser illumination under applied bias, and normalized the measured time-domain signal by the slope of the switch’s response curve.\cite{McIver2020,Kipp2024} In Figure~A-2(b) in the supplementary material we show the response of the d-PCS on each side. A comparison of the left and right sides after calibration is shown in \prettyref{fig:4}(a). The d-PCSs detected nearly identical signals, indicating the high reproducibility and reliability of the nanofabrication process. Incorporating two symmetrical branches into a single circuit also allows simultaneous measurement of the reference and signal under identical experimental conditions, significantly reducing the total measurement time.

Monolithic fabrication with $\alpha$-Si not only provides effective and reliable device performance, but also yields PCSs with a high breakdown voltage. Notably, the generated THz field scales linearly with the bias, as can be seen in \prettyref{fig:4}(b). Applying fields up to at least \SI{200}{kV/cm} does not lead to visible damage nor decrease of the signal, while achieving THz fields on the order of \SI{1}{kV/cm}.  Circuits with LT-GaAs as PCS are reported to achieve THz field amplitudes of up to {40}~{kV/cm}\cite{Diaz2023}, however, it has a break down field of {94}~{kV/cm}\cite{Pottsphd2023}. The high breakdown voltage of $\alpha$-Si, in comparison, allows for flexibility in circuitry design. \\

The durability of the silicon PCS, in combination with the AC-coupled design, can be advantageous for measuring sensitive samples, as the generation and detection are galvanically isolated.
Moreover, a sample under investigation would be placed underneath and between the metal traces of the CPS, as indicated in \prettyref{fig:1}. For control of carrier densities in electronic materials and devices, it is common to use electrostatic gates. Using our circuitry would allow for a sample design which uses the two metal traces as gates. The galvanic isolation in our circuits enables independent control of THz generation, detection and the application of static potentials, simplifying the device integration with our circuitry.

\section[floatfix]{\label{sec:level3}Conclusion}
We have developed an improved optoelectronic circuit architecture for on-chip THz applications, with a particular focus on enabling high-fidelity time-domain THz spectroscopy. The design features a THz field generator and two detectors for sample measurement and referencing, all fabricated monolithically from electron-beam evaporated $\alpha$-Si. Compared to conventional DC-coupled designs, our circuit exhibits pure odd-mode propagation, an expanded operational bandwidth, and enhanced THz field strength, in comparison to the DC-coupled design.

These advances establish a strong foundation for linear THz spectroscopy, where pure mode propagation is critical for preserving in-plane field information and avoiding signal loss. Moving forward, the circuit is also well suited for nonlinear THz spectroscopy, where mixed-mode propagation could complicate the interpretation of higher-order responses. By supporting pure-mode operation, our design simplifies the analysis of nonlinear conductivities and higher-order optical processes.

Furthermore, the reproducibility of the monolithic fabrication, combined with the integrated \textit{in situ} referencing and galvanic isolation between generation and detection, enables precise field calibration and device integration. This architecture not only improves spectroscopy workflows but also offers a robust platform for broader applications in THz electronics, including cavity engineering, stripline gating, and the study of quantum materials, biological systems, and nanoscale devices.

\section*{Supplementary material}
See supplementary material for time evolution of the THz fields, relation of THz field to antenna-arm length, source data for THz field dependence on generator bias, raw data of the signals of two sides within one device and their calibration curves, and our assessment of the quasi-TEM mode in our circuits. 

\begin{acknowledgments}
We acknowledge the support of B. Fiedler, B. Höhling, E. König, M. Barthelmess, E. Wang, J. Adelinia, M. Chavez-Cervantes for assistance in fabrication of electronic devices and support in the clean room. 
\end{acknowledgments}

\section*{Author Declaration}
\subsubsection*{Conflict of interest}
The authors have no conflicts to disclose.
\subsection*{Author Contribution}

J.W.M. conceived the experiment and supervised the project. K.K., B.S. and G.K. fabricated the devices. B.S. designed and built the measurement setup with technical upgrades by M.W.D. and K.K.. K.K. and B.S. performed measurements and analyzed the data. B.S., H.M.B. and M.W.D. developed the data analysis framework. Custom measurement electronics and circuit simulations were provided by T.M. and G.M.. K.K.,  A.M.P. and J.W.M. wrote the manuscript with input from all authors.

\section*{Funding}
We acknowledge support from the Max Planck-New York City Center for Non-Equilibrium Quantum Phenomena. G.K. acknowledges support by the German Research Foundation through the Cluster of Excellence CUI: Advanced Imaging of Matter (EXC 2056, project ID 390715994). H.M.B. and M.W.D. acknowledge support from the Alexander von Humboldt Foundation. H.M.B. acknowledges financial support from the European Union under the Marie Sklodowska-Curie Grant Agreement no. 101062921 (Twist-TOC). The fabrication of THz circuit optimization devices was supported by the U.S. Department of Energy, Office of Science, Basic Energy Sciences, under Early Career Award DE-SC0024334. At the early stages of the project, K.K. and G.K. were supported by SFB 925 - project 170620586/Deutsche Forschungsgemeinschaft (German Research Foundation).

\section*{Data Availability Statement}

The data that support the findings of this study are available from the corresponding author upon reasonable request.

\section*{References}
\bibliography{aipsamp}

\providecommand{\noopsort}[1]{}\providecommand{\singleletter}[1]{#1}%
\begin{thebibliography}{40}%
\makeatletter
\providecommand \@ifxundefined [1]{%
 \@ifx{#1\undefined}
}%
\providecommand \@ifnum [1]{%
 \ifnum #1\expandafter \@firstoftwo
 \else \expandafter \@secondoftwo
 \fi
}%
\providecommand \@ifx [1]{%
 \ifx #1\expandafter \@firstoftwo
 \else \expandafter \@secondoftwo
 \fi
}%
\providecommand \natexlab [1]{#1}%
\providecommand \enquote  [1]{``#1''}%
\providecommand \bibnamefont  [1]{#1}%
\providecommand \bibfnamefont [1]{#1}%
\providecommand \citenamefont [1]{#1}%
\providecommand \href@noop [0]{\@secondoftwo}%
\providecommand \href [0]{\begingroup \@sanitize@url \@href}%
\providecommand \@href[1]{\@@startlink{#1}\@@href}%
\providecommand \@@href[1]{\endgroup#1\@@endlink}%
\providecommand \@sanitize@url [0]{\catcode `\\12\catcode `\$12\catcode `\&12\catcode `\#12\catcode `\^12\catcode `\_12\catcode `\%12\relax}%
\providecommand \@@startlink[1]{}%
\providecommand \@@endlink[0]{}%
\providecommand \url  [0]{\begingroup\@sanitize@url \@url }%
\providecommand \@url [1]{\endgroup\@href {#1}{\urlprefix }}%
\providecommand \urlprefix  [0]{URL }%
\providecommand \Eprint [0]{\href }%
\providecommand \doibase [0]{http://dx.doi.org/}%
\providecommand \selectlanguage [0]{\@gobble}%
\providecommand \bibinfo  [0]{\@secondoftwo}%
\providecommand \bibfield  [0]{\@secondoftwo}%
\providecommand \translation [1]{[#1]}%
\providecommand \BibitemOpen [0]{}%
\providecommand \bibitemStop [0]{}%
\providecommand \bibitemNoStop [0]{.\EOS\space}%
\providecommand \EOS [0]{\spacefactor3000\relax}%
\providecommand \BibitemShut  [1]{\csname bibitem#1\endcsname}%
\let\auto@bib@innerbib\@empty
\bibitem [{\citenamefont {{Ishigaki}}\ \emph {et~al.}(2012)\citenamefont {{Ishigaki}}, \citenamefont {{Shiraishi}}, \citenamefont {{Suzuki}}, \citenamefont {{Asada}}, \citenamefont {{Nishiyama}},\ and\ \citenamefont {{Arai}}}]{Ishigaki2012}%
  \BibitemOpen
  \bibfield  {author} {\bibinfo {author} {\bibfnamefont {K.}~\bibnamefont {{Ishigaki}}}, \bibinfo {author} {\bibfnamefont {M.}~\bibnamefont {{Shiraishi}}}, \bibinfo {author} {\bibfnamefont {S.}~\bibnamefont {{Suzuki}}}, \bibinfo {author} {\bibfnamefont {M.}~\bibnamefont {{Asada}}}, \bibinfo {author} {\bibfnamefont {N.}~\bibnamefont {{Nishiyama}}}, \ and\ \bibinfo {author} {\bibfnamefont {S.}~\bibnamefont {{Arai}}},\ }\bibfield  {title} {\enquote {\bibinfo {title} {{Direct intensity modulation and wireless data transmission characteristics of terahertz-oscillating resonant tunnelling diodes}},}\ }\href {\doibase 10.1049/el.2012.0849} {\bibfield  {journal} {\bibinfo  {journal} {Electronics Letters}\ }\textbf {\bibinfo {volume} {48}},\ \bibinfo {pages} {582} (\bibinfo {year} {2012})}\BibitemShut {NoStop}%
\bibitem [{\citenamefont {{Rappaport}}\ \emph {et~al.}(2019)\citenamefont {{Rappaport}}, \citenamefont {{Xing}}, \citenamefont {{Kanhere}}, \citenamefont {{Ju}}, \citenamefont {{Madanayake}}, \citenamefont {{Mandal}}, \citenamefont {{Alkhateeb}},\ and\ \citenamefont {{Trichopoulos}}}]{Rappaport2019}%
  \BibitemOpen
  \bibfield  {author} {\bibinfo {author} {\bibfnamefont {T.~S.}\ \bibnamefont {{Rappaport}}}, \bibinfo {author} {\bibfnamefont {Y.}~\bibnamefont {{Xing}}}, \bibinfo {author} {\bibfnamefont {O.}~\bibnamefont {{Kanhere}}}, \bibinfo {author} {\bibfnamefont {S.}~\bibnamefont {{Ju}}}, \bibinfo {author} {\bibfnamefont {A.}~\bibnamefont {{Madanayake}}}, \bibinfo {author} {\bibfnamefont {S.}~\bibnamefont {{Mandal}}}, \bibinfo {author} {\bibfnamefont {A.}~\bibnamefont {{Alkhateeb}}}, \ and\ \bibinfo {author} {\bibfnamefont {G.~C.}\ \bibnamefont {{Trichopoulos}}},\ }\bibfield  {title} {\enquote {\bibinfo {title} {{Wireless Communications and Applications Above 100 GHz: Opportunities and Challenges for 6G and Beyond}},}\ }\href {\doibase 10.1109/ACCESS.2019.2921522} {\bibfield  {journal} {\bibinfo  {journal} {IEEE Access}\ }\textbf {\bibinfo {volume} {7}},\ \bibinfo {pages} {78729--78757} (\bibinfo {year} {2019})}\BibitemShut {NoStop}%
\bibitem [{\citenamefont {Mittleman}\ \emph {et~al.}(1998)\citenamefont {Mittleman}, \citenamefont {Jacobsen}, \citenamefont {Neelamani}, \citenamefont {Baraniuk},\ and\ \citenamefont {Nuss}}]{Mittelmann1998}%
  \BibitemOpen
  \bibfield  {author} {\bibinfo {author} {\bibfnamefont {D.}~\bibnamefont {Mittleman}}, \bibinfo {author} {\bibfnamefont {R.}~\bibnamefont {Jacobsen}}, \bibinfo {author} {\bibfnamefont {R.}~\bibnamefont {Neelamani}}, \bibinfo {author} {\bibfnamefont {R.}~\bibnamefont {Baraniuk}}, \ and\ \bibinfo {author} {\bibfnamefont {M.}~\bibnamefont {Nuss}},\ }\bibfield  {title} {\enquote {\bibinfo {title} {{Gas sensing using terahertz time-domain spectroscopy}},}\ }\href {\doibase 10.1007/s003400050520} {\bibfield  {journal} {\bibinfo  {journal} {Applied Physics B}\ }\textbf {\bibinfo {volume} {67}},\ \bibinfo {pages} {379--390} (\bibinfo {year} {1998})}\BibitemShut {NoStop}%
\bibitem [{\citenamefont {Weide}, \citenamefont {Murakowski},\ and\ \citenamefont {Keilmann}(2000)}]{Weide2000}%
  \BibitemOpen
  \bibfield  {author} {\bibinfo {author} {\bibfnamefont {D.~V.~d.}\ \bibnamefont {Weide}}, \bibinfo {author} {\bibfnamefont {J.}~\bibnamefont {Murakowski}}, \ and\ \bibinfo {author} {\bibfnamefont {F.}~\bibnamefont {Keilmann}},\ }\bibfield  {title} {\enquote {\bibinfo {title} {{Gas-absorption spectroscopy with electronic terahertz techniques}},}\ }\href {\doibase 10.1109/22.841967} {\bibfield  {journal} {\bibinfo  {journal} {IEEE Transactions on Microwave Theory and Techniques}\ }\textbf {\bibinfo {volume} {48}},\ \bibinfo {pages} {740--743} (\bibinfo {year} {2000})}\BibitemShut {NoStop}%
\bibitem [{\citenamefont {Federici}\ \emph {et~al.}(2005)\citenamefont {Federici}, \citenamefont {Schulkin}, \citenamefont {Huang}, \citenamefont {Gary}, \citenamefont {Barat}, \citenamefont {Oliveira},\ and\ \citenamefont {Zimdars}}]{Federici1005}%
  \BibitemOpen
  \bibfield  {author} {\bibinfo {author} {\bibfnamefont {J.~F.}\ \bibnamefont {Federici}}, \bibinfo {author} {\bibfnamefont {B.}~\bibnamefont {Schulkin}}, \bibinfo {author} {\bibfnamefont {F.}~\bibnamefont {Huang}}, \bibinfo {author} {\bibfnamefont {D.}~\bibnamefont {Gary}}, \bibinfo {author} {\bibfnamefont {R.}~\bibnamefont {Barat}}, \bibinfo {author} {\bibfnamefont {F.}~\bibnamefont {Oliveira}}, \ and\ \bibinfo {author} {\bibfnamefont {D.}~\bibnamefont {Zimdars}},\ }\bibfield  {title} {\enquote {\bibinfo {title} {{THz imaging and sensing for security applications—explosives, weapons and drugs}},}\ }\href {\doibase 10.1088/0268-1242/20/7/018} {\bibfield  {journal} {\bibinfo  {journal} {Semiconductor Science and Technology}\ }\textbf {\bibinfo {volume} {20}},\ \bibinfo {pages} {S266} (\bibinfo {year} {2005})}\BibitemShut {NoStop}%
\bibitem [{\citenamefont {Nagel}, \citenamefont {Först},\ and\ \citenamefont {Kurz}(2006)}]{Nagel2006}%
  \BibitemOpen
  \bibfield  {author} {\bibinfo {author} {\bibfnamefont {M.}~\bibnamefont {Nagel}}, \bibinfo {author} {\bibfnamefont {M.}~\bibnamefont {Först}}, \ and\ \bibinfo {author} {\bibfnamefont {H.}~\bibnamefont {Kurz}},\ }\bibfield  {title} {\enquote {\bibinfo {title} {{THz biosensing devices: fundamentals and technology}},}\ }\href {\doibase 10.1088/0953-8984/18/18/s07} {\bibfield  {journal} {\bibinfo  {journal} {Journal of Physics: Condensed Matter}\ }\textbf {\bibinfo {volume} {18}},\ \bibinfo {pages} {S601} (\bibinfo {year} {2006})}\BibitemShut {NoStop}%
\bibitem [{\citenamefont {Siegel}(2004)}]{Siegel2004}%
  \BibitemOpen
  \bibfield  {author} {\bibinfo {author} {\bibfnamefont {P.~H.}\ \bibnamefont {Siegel}},\ }\bibfield  {title} {\enquote {\bibinfo {title} {{Terahertz Technology in Biology and Medicine}},}\ }\href {\doibase 10.1109/tmtt.2004.835916} {\bibfield  {journal} {\bibinfo  {journal} {IEEE Transactions on Microwave Theory and Techniques}\ }\textbf {\bibinfo {volume} {52}},\ \bibinfo {pages} {2438--2447} (\bibinfo {year} {2004})}\BibitemShut {NoStop}%
\bibitem [{\citenamefont {Woodward}\ \emph {et~al.}(2003)\citenamefont {Woodward}, \citenamefont {Wallace}, \citenamefont {Arnone}, \citenamefont {Linfield},\ and\ \citenamefont {Pepper}}]{Woodward2003}%
  \BibitemOpen
  \bibfield  {author} {\bibinfo {author} {\bibfnamefont {R.}~\bibnamefont {Woodward}}, \bibinfo {author} {\bibfnamefont {V.}~\bibnamefont {Wallace}}, \bibinfo {author} {\bibfnamefont {D.}~\bibnamefont {Arnone}}, \bibinfo {author} {\bibfnamefont {E.}~\bibnamefont {Linfield}}, \ and\ \bibinfo {author} {\bibfnamefont {M.}~\bibnamefont {Pepper}},\ }\bibfield  {title} {\enquote {\bibinfo {title} {{Terahertz Pulsed Imaging of Skin Cancer in the Time and Frequency Domain}},}\ }\href {\doibase 10.1023/a:1024409329416} {\bibfield  {journal} {\bibinfo  {journal} {Journal of Biological Physics}\ }\textbf {\bibinfo {volume} {29}},\ \bibinfo {pages} {257--259} (\bibinfo {year} {2003})}\BibitemShut {NoStop}%
\bibitem [{\citenamefont {Basov}\ \emph {et~al.}(2011)\citenamefont {Basov}, \citenamefont {Averitt}, \citenamefont {Marel}, \citenamefont {Dressel},\ and\ \citenamefont {Haule}}]{Basov2011}%
  \BibitemOpen
  \bibfield  {author} {\bibinfo {author} {\bibfnamefont {D.~N.}\ \bibnamefont {Basov}}, \bibinfo {author} {\bibfnamefont {R.~D.}\ \bibnamefont {Averitt}}, \bibinfo {author} {\bibfnamefont {D.~v.~d.}\ \bibnamefont {Marel}}, \bibinfo {author} {\bibfnamefont {M.}~\bibnamefont {Dressel}}, \ and\ \bibinfo {author} {\bibfnamefont {K.}~\bibnamefont {Haule}},\ }\bibfield  {title} {\enquote {\bibinfo {title} {{Electrodynamics of correlated electron materials}},}\ }\href {\doibase 10.1103/revmodphys.83.471} {\bibfield  {journal} {\bibinfo  {journal} {Reviews of Modern Physics}\ }\textbf {\bibinfo {volume} {83}},\ \bibinfo {pages} {471--541} (\bibinfo {year} {2011})},\ \Eprint {http://arxiv.org/abs/1106.2309} {1106.2309} \BibitemShut {NoStop}%
\bibitem [{\citenamefont {Basov}, \citenamefont {Averitt},\ and\ \citenamefont {Hsieh}(2017)}]{Basov2017}%
  \BibitemOpen
  \bibfield  {author} {\bibinfo {author} {\bibfnamefont {D.~N.}\ \bibnamefont {Basov}}, \bibinfo {author} {\bibfnamefont {R.~D.}\ \bibnamefont {Averitt}}, \ and\ \bibinfo {author} {\bibfnamefont {D.}~\bibnamefont {Hsieh}},\ }\bibfield  {title} {\enquote {\bibinfo {title} {{Towards properties on demand in quantum materials}},}\ }\href {\doibase 10.1038/nmat5017} {\bibfield  {journal} {\bibinfo  {journal} {Nature Materials}\ }\textbf {\bibinfo {volume} {16}},\ \bibinfo {pages} {1077--1088} (\bibinfo {year} {2017})}\BibitemShut {NoStop}%
\bibitem [{\citenamefont {Torre}\ \emph {et~al.}(2021)\citenamefont {Torre}, \citenamefont {Kennes}, \citenamefont {Claassen}, \citenamefont {Gerber}, \citenamefont {McIver},\ and\ \citenamefont {Sentef}}]{Torre2021}%
  \BibitemOpen
  \bibfield  {author} {\bibinfo {author} {\bibfnamefont {A.~d.~l.}\ \bibnamefont {Torre}}, \bibinfo {author} {\bibfnamefont {D.~M.}\ \bibnamefont {Kennes}}, \bibinfo {author} {\bibfnamefont {M.}~\bibnamefont {Claassen}}, \bibinfo {author} {\bibfnamefont {S.}~\bibnamefont {Gerber}}, \bibinfo {author} {\bibfnamefont {J.~W.}\ \bibnamefont {McIver}}, \ and\ \bibinfo {author} {\bibfnamefont {M.~A.}\ \bibnamefont {Sentef}},\ }\bibfield  {title} {\enquote {\bibinfo {title} {{Colloquium: Nonthermal pathways to ultrafast control in quantum materials}},}\ }\href {\doibase 10.1103/revmodphys.93.041002} {\bibfield  {journal} {\bibinfo  {journal} {Reviews of Modern Physics}\ }\textbf {\bibinfo {volume} {93}},\ \bibinfo {pages} {041002} (\bibinfo {year} {2021})},\ \Eprint {http://arxiv.org/abs/2103.14888} {2103.14888} \BibitemShut {NoStop}%
\bibitem [{\citenamefont {Auston}(1975)}]{Auston1975}%
  \BibitemOpen
  \bibfield  {author} {\bibinfo {author} {\bibfnamefont {D.~H.}\ \bibnamefont {Auston}},\ }\bibfield  {title} {\enquote {\bibinfo {title} {{Picosecond optoelectronic switching and gating in silicon}},}\ }\href {\doibase 10.1063/1.88079} {\bibfield  {journal} {\bibinfo  {journal} {Applied Physics Letters}\ }\textbf {\bibinfo {volume} {26}},\ \bibinfo {pages} {101--103} (\bibinfo {year} {1975})}\BibitemShut {NoStop}%
\bibitem [{\citenamefont {Gallagher}\ \emph {et~al.}(2019)\citenamefont {Gallagher}, \citenamefont {Yang}, \citenamefont {Lyu}, \citenamefont {Tian}, \citenamefont {Kou}, \citenamefont {Zhang}, \citenamefont {Watanabe}, \citenamefont {Taniguchi},\ and\ \citenamefont {Wang}}]{Gallagher2019}%
  \BibitemOpen
  \bibfield  {author} {\bibinfo {author} {\bibfnamefont {P.}~\bibnamefont {Gallagher}}, \bibinfo {author} {\bibfnamefont {C.-S.}\ \bibnamefont {Yang}}, \bibinfo {author} {\bibfnamefont {T.}~\bibnamefont {Lyu}}, \bibinfo {author} {\bibfnamefont {F.}~\bibnamefont {Tian}}, \bibinfo {author} {\bibfnamefont {R.}~\bibnamefont {Kou}}, \bibinfo {author} {\bibfnamefont {H.}~\bibnamefont {Zhang}}, \bibinfo {author} {\bibfnamefont {K.}~\bibnamefont {Watanabe}}, \bibinfo {author} {\bibfnamefont {T.}~\bibnamefont {Taniguchi}}, \ and\ \bibinfo {author} {\bibfnamefont {F.}~\bibnamefont {Wang}},\ }\bibfield  {title} {\enquote {\bibinfo {title} {{Quantum-critical conductivity of the Dirac fluid in graphene}},}\ }\href {\doibase 10.1126/science.aat8687} {\bibfield  {journal} {\bibinfo  {journal} {Science}\ }\textbf {\bibinfo {volume} {364}},\ \bibinfo {pages} {158--162} (\bibinfo {year} {2019})}\BibitemShut {NoStop}%
\bibitem [{\citenamefont {Zhao}\ \emph {et~al.}(2023)\citenamefont {Zhao}, \citenamefont {Wang}, \citenamefont {Chen}, \citenamefont {Zhang}, \citenamefont {Watanabe}, \citenamefont {Taniguchi}, \citenamefont {Zettl},\ and\ \citenamefont {Wang}}]{Zhao2023}%
  \BibitemOpen
  \bibfield  {author} {\bibinfo {author} {\bibfnamefont {W.}~\bibnamefont {Zhao}}, \bibinfo {author} {\bibfnamefont {S.}~\bibnamefont {Wang}}, \bibinfo {author} {\bibfnamefont {S.}~\bibnamefont {Chen}}, \bibinfo {author} {\bibfnamefont {Z.}~\bibnamefont {Zhang}}, \bibinfo {author} {\bibfnamefont {K.}~\bibnamefont {Watanabe}}, \bibinfo {author} {\bibfnamefont {T.}~\bibnamefont {Taniguchi}}, \bibinfo {author} {\bibfnamefont {A.}~\bibnamefont {Zettl}}, \ and\ \bibinfo {author} {\bibfnamefont {F.}~\bibnamefont {Wang}},\ }\bibfield  {title} {\enquote {\bibinfo {title} {{Observation of hydrodynamic plasmons and energy waves in graphene}},}\ }\href {\doibase 10.1038/s41586-022-05619-8} {\bibfield  {journal} {\bibinfo  {journal} {Nature}\ }\textbf {\bibinfo {volume} {614}},\ \bibinfo {pages} {688--693} (\bibinfo {year} {2023})}\BibitemShut {NoStop}%
\bibitem [{\citenamefont {Smith}\ and\ \citenamefont {Darcie}(2019)}]{Smith2019}%
  \BibitemOpen
  \bibfield  {author} {\bibinfo {author} {\bibfnamefont {R.}~\bibnamefont {Smith}}\ and\ \bibinfo {author} {\bibfnamefont {T.}~\bibnamefont {Darcie}},\ }\bibfield  {title} {\enquote {\bibinfo {title} {{Demonstration of a low-distortion terahertz system-on-chip using a CPS waveguide on a thin membrane substrate}},}\ }\href {\doibase 10.1364/oe.27.013653} {\bibfield  {journal} {\bibinfo  {journal} {Optics Express}\ }\textbf {\bibinfo {volume} {27}},\ \bibinfo {pages} {13653} (\bibinfo {year} {2019})}\BibitemShut {NoStop}%
\bibitem [{\citenamefont {Potts}\ \emph {et~al.}(2023)\citenamefont {Potts}, \citenamefont {Nayak}, \citenamefont {Nagel}, \citenamefont {Kaj}, \citenamefont {Stamenic}, \citenamefont {John}, \citenamefont {Averitt},\ and\ \citenamefont {Young}}]{Potts2023}%
  \BibitemOpen
  \bibfield  {author} {\bibinfo {author} {\bibfnamefont {A.~M.}\ \bibnamefont {Potts}}, \bibinfo {author} {\bibfnamefont {A.~K.}\ \bibnamefont {Nayak}}, \bibinfo {author} {\bibfnamefont {M.}~\bibnamefont {Nagel}}, \bibinfo {author} {\bibfnamefont {K.}~\bibnamefont {Kaj}}, \bibinfo {author} {\bibfnamefont {B.}~\bibnamefont {Stamenic}}, \bibinfo {author} {\bibfnamefont {D.~D.}\ \bibnamefont {John}}, \bibinfo {author} {\bibfnamefont {R.~D.}\ \bibnamefont {Averitt}}, \ and\ \bibinfo {author} {\bibfnamefont {A.~F.}\ \bibnamefont {Young}},\ }\bibfield  {title} {\enquote {\bibinfo {title} {{On-Chip Time-Domain Terahertz Spectroscopy of Superconducting Films below the Diffraction Limit}},}\ }\href {\doibase 10.1021/acs.nanolett.3c00412} {\bibfield  {journal} {\bibinfo  {journal} {Nano Letters}\ }\textbf {\bibinfo {volume} {23}},\ \bibinfo {pages} {3835--3841} (\bibinfo {year} {2023})},\ \Eprint {http://arxiv.org/abs/2302.05434} {2302.05434} \BibitemShut {NoStop}%
\bibitem [{\citenamefont {Yoshioka}\ \emph {et~al.}(2022)\citenamefont {Yoshioka}, \citenamefont {Wakamura}, \citenamefont {Hashisaka}, \citenamefont {Watanabe}, \citenamefont {Taniguchi},\ and\ \citenamefont {Kumada}}]{Yoshioka2022}%
  \BibitemOpen
  \bibfield  {author} {\bibinfo {author} {\bibfnamefont {K.}~\bibnamefont {Yoshioka}}, \bibinfo {author} {\bibfnamefont {T.}~\bibnamefont {Wakamura}}, \bibinfo {author} {\bibfnamefont {M.}~\bibnamefont {Hashisaka}}, \bibinfo {author} {\bibfnamefont {K.}~\bibnamefont {Watanabe}}, \bibinfo {author} {\bibfnamefont {T.}~\bibnamefont {Taniguchi}}, \ and\ \bibinfo {author} {\bibfnamefont {N.}~\bibnamefont {Kumada}},\ }\bibfield  {title} {\enquote {\bibinfo {title} {{Ultrafast intrinsic optical-to-electrical conversion dynamics in a graphene photodetector}},}\ }\href {\doibase 10.1038/s41566-022-01058-z} {\bibfield  {journal} {\bibinfo  {journal} {Nature Photonics}\ }\textbf {\bibinfo {volume} {16}},\ \bibinfo {pages} {718--723} (\bibinfo {year} {2022})},\ \Eprint {http://arxiv.org/abs/2203.05752} {2203.05752} \BibitemShut {NoStop}%
\bibitem [{\citenamefont {Yoshioka}\ \emph {et~al.}(2024)\citenamefont {Yoshioka}, \citenamefont {Bernard}, \citenamefont {Wakamura}, \citenamefont {Hashisaka}, \citenamefont {Sasaki}, \citenamefont {Sasaki}, \citenamefont {Watanabe}, \citenamefont {Taniguchi},\ and\ \citenamefont {Kumada}}]{Yoshioka2024}%
  \BibitemOpen
  \bibfield  {author} {\bibinfo {author} {\bibfnamefont {K.}~\bibnamefont {Yoshioka}}, \bibinfo {author} {\bibfnamefont {G.}~\bibnamefont {Bernard}}, \bibinfo {author} {\bibfnamefont {T.}~\bibnamefont {Wakamura}}, \bibinfo {author} {\bibfnamefont {M.}~\bibnamefont {Hashisaka}}, \bibinfo {author} {\bibfnamefont {K.-i.}\ \bibnamefont {Sasaki}}, \bibinfo {author} {\bibfnamefont {S.}~\bibnamefont {Sasaki}}, \bibinfo {author} {\bibfnamefont {K.}~\bibnamefont {Watanabe}}, \bibinfo {author} {\bibfnamefont {T.}~\bibnamefont {Taniguchi}}, \ and\ \bibinfo {author} {\bibfnamefont {N.}~\bibnamefont {Kumada}},\ }\bibfield  {title} {\enquote {\bibinfo {title} {{On-chip transfer of ultrashort graphene plasmon wave packets using terahertz electronics}},}\ }\href {\doibase 10.1038/s41928-024-01197-x} {\bibfield  {journal} {\bibinfo  {journal} {Nature Electronics}\ ,\ \bibinfo {pages} {1--8}} (\bibinfo {year} {2024})}\BibitemShut {NoStop}%
\bibitem [{\citenamefont {Seo}\ \emph {et~al.}(2024)\citenamefont {Seo}, \citenamefont {Lu}, \citenamefont {Park}, \citenamefont {Yang}, \citenamefont {Xia}, \citenamefont {Ye}, \citenamefont {Yao}, \citenamefont {Han}, \citenamefont {Shi}, \citenamefont {Watanabe}, \citenamefont {Taniguchi}, \citenamefont {Yacoby},\ and\ \citenamefont {Ju}}]{Seo2024}%
  \BibitemOpen
  \bibfield  {author} {\bibinfo {author} {\bibfnamefont {J.}~\bibnamefont {Seo}}, \bibinfo {author} {\bibfnamefont {Z.}~\bibnamefont {Lu}}, \bibinfo {author} {\bibfnamefont {S.}~\bibnamefont {Park}}, \bibinfo {author} {\bibfnamefont {J.}~\bibnamefont {Yang}}, \bibinfo {author} {\bibfnamefont {F.}~\bibnamefont {Xia}}, \bibinfo {author} {\bibfnamefont {S.}~\bibnamefont {Ye}}, \bibinfo {author} {\bibfnamefont {Y.}~\bibnamefont {Yao}}, \bibinfo {author} {\bibfnamefont {T.}~\bibnamefont {Han}}, \bibinfo {author} {\bibfnamefont {L.}~\bibnamefont {Shi}}, \bibinfo {author} {\bibfnamefont {K.}~\bibnamefont {Watanabe}}, \bibinfo {author} {\bibfnamefont {T.}~\bibnamefont {Taniguchi}}, \bibinfo {author} {\bibfnamefont {A.}~\bibnamefont {Yacoby}}, \ and\ \bibinfo {author} {\bibfnamefont {L.}~\bibnamefont {Ju}},\ }\bibfield  {title} {\enquote {\bibinfo {title} {{On-Chip Terahertz Spectroscopy for Dual-Gated van der Waals Heterostructures at Cryogenic Temperatures}},}\ }\href {\doibase 10.1021/acs.nanolett.4c04137}
  {\bibfield  {journal} {\bibinfo  {journal} {Nano Letters}\ }\textbf {\bibinfo {volume} {24}},\ \bibinfo {pages} {15060--15067} (\bibinfo {year} {2024})},\ \Eprint {http://arxiv.org/abs/2409.19726} {2409.19726} \BibitemShut {NoStop}%
\bibitem [{\citenamefont {Kipp}\ \emph {et~al.}(2024)\citenamefont {Kipp}, \citenamefont {Bretscher}, \citenamefont {Schulte}, \citenamefont {Herrmann}, \citenamefont {Kusyak}, \citenamefont {Day}, \citenamefont {Kesavan}, \citenamefont {Matsuyama}, \citenamefont {Li}, \citenamefont {Langner}, \citenamefont {Hagelstein}, \citenamefont {Sturm}, \citenamefont {Potts}, \citenamefont {Eckhardt}, \citenamefont {Huang}, \citenamefont {Watanabe}, \citenamefont {Taniguchi}, \citenamefont {Rubio}, \citenamefont {Kennes}, \citenamefont {Sentef}, \citenamefont {Baudin}, \citenamefont {Meier}, \citenamefont {Michael},\ and\ \citenamefont {McIver}}]{Kipp2024}%
  \BibitemOpen
  \bibfield  {author} {\bibinfo {author} {\bibfnamefont {G.}~\bibnamefont {Kipp}}, \bibinfo {author} {\bibfnamefont {H.~M.}\ \bibnamefont {Bretscher}}, \bibinfo {author} {\bibfnamefont {B.}~\bibnamefont {Schulte}}, \bibinfo {author} {\bibfnamefont {D.}~\bibnamefont {Herrmann}}, \bibinfo {author} {\bibfnamefont {K.}~\bibnamefont {Kusyak}}, \bibinfo {author} {\bibfnamefont {M.~W.}\ \bibnamefont {Day}}, \bibinfo {author} {\bibfnamefont {S.}~\bibnamefont {Kesavan}}, \bibinfo {author} {\bibfnamefont {T.}~\bibnamefont {Matsuyama}}, \bibinfo {author} {\bibfnamefont {X.}~\bibnamefont {Li}}, \bibinfo {author} {\bibfnamefont {S.~M.}\ \bibnamefont {Langner}}, \bibinfo {author} {\bibfnamefont {J.}~\bibnamefont {Hagelstein}}, \bibinfo {author} {\bibfnamefont {F.}~\bibnamefont {Sturm}}, \bibinfo {author} {\bibfnamefont {A.~M.}\ \bibnamefont {Potts}}, \bibinfo {author} {\bibfnamefont {C.~J.}\ \bibnamefont {Eckhardt}}, \bibinfo {author} {\bibfnamefont {Y.}~\bibnamefont {Huang}}, \bibinfo {author} {\bibfnamefont
  {K.}~\bibnamefont {Watanabe}}, \bibinfo {author} {\bibfnamefont {T.}~\bibnamefont {Taniguchi}}, \bibinfo {author} {\bibfnamefont {A.}~\bibnamefont {Rubio}}, \bibinfo {author} {\bibfnamefont {D.~M.}\ \bibnamefont {Kennes}}, \bibinfo {author} {\bibfnamefont {M.~A.}\ \bibnamefont {Sentef}}, \bibinfo {author} {\bibfnamefont {E.}~\bibnamefont {Baudin}}, \bibinfo {author} {\bibfnamefont {G.}~\bibnamefont {Meier}}, \bibinfo {author} {\bibfnamefont {M.~H.}\ \bibnamefont {Michael}}, \ and\ \bibinfo {author} {\bibfnamefont {J.~W.}\ \bibnamefont {McIver}},\ }\bibfield  {title} {\enquote {\bibinfo {title} {{Cavity electrodynamics of van der Waals heterostructures}},}\ }\href@noop {} {\bibfield  {journal} {\bibinfo  {journal} {arXiv}\ } (\bibinfo {year} {2024})},\ \Eprint {http://arxiv.org/abs/2403.19745} {2403.19745} \BibitemShut {NoStop}%
\bibitem [{\citenamefont {Karnetzky}\ \emph {et~al.}(2018)\citenamefont {Karnetzky}, \citenamefont {Zimmermann}, \citenamefont {Trummer}, \citenamefont {Sierra}, \citenamefont {Wörle}, \citenamefont {Kienberger},\ and\ \citenamefont {Holleitner}}]{Karnetzky2018}%
  \BibitemOpen
  \bibfield  {author} {\bibinfo {author} {\bibfnamefont {C.}~\bibnamefont {Karnetzky}}, \bibinfo {author} {\bibfnamefont {P.}~\bibnamefont {Zimmermann}}, \bibinfo {author} {\bibfnamefont {C.}~\bibnamefont {Trummer}}, \bibinfo {author} {\bibfnamefont {C.~D.}\ \bibnamefont {Sierra}}, \bibinfo {author} {\bibfnamefont {M.}~\bibnamefont {Wörle}}, \bibinfo {author} {\bibfnamefont {R.}~\bibnamefont {Kienberger}}, \ and\ \bibinfo {author} {\bibfnamefont {A.}~\bibnamefont {Holleitner}},\ }\bibfield  {title} {\enquote {\bibinfo {title} {{Towards femtosecond on-chip electronics based on plasmonic hot electron nano-emitters}},}\ }\href {\doibase 10.1038/s41467-018-04666-y} {\bibfield  {journal} {\bibinfo  {journal} {Nature Communications}\ }\textbf {\bibinfo {volume} {9}},\ \bibinfo {pages} {2471} (\bibinfo {year} {2018})}\BibitemShut {NoStop}%
\bibitem [{\citenamefont {Wu}\ \emph {et~al.}(2015)\citenamefont {Wu}, \citenamefont {{Alexander S. Mayorov}}, \citenamefont {{Christopher D. Wood}}, \citenamefont {{Divyang Mistry}}, \citenamefont {{Lianhe Li}}, \citenamefont {{Wilson Muchenje}}, \citenamefont {{Mark C. Rosamond}}, \citenamefont {{Li Chen}}, \citenamefont {{Edmund H. Linfield}}, \citenamefont {{A. Giles Davies}},\ and\ \citenamefont {{John E. Cunningham}}}]{Wu2015}%
  \BibitemOpen
  \bibfield  {author} {\bibinfo {author} {\bibfnamefont {J.}~\bibnamefont {Wu}}, \bibinfo {author} {\bibnamefont {{Alexander S. Mayorov}}}, \bibinfo {author} {\bibnamefont {{Christopher D. Wood}}}, \bibinfo {author} {\bibnamefont {{Divyang Mistry}}}, \bibinfo {author} {\bibnamefont {{Lianhe Li}}}, \bibinfo {author} {\bibnamefont {{Wilson Muchenje}}}, \bibinfo {author} {\bibnamefont {{Mark C. Rosamond}}}, \bibinfo {author} {\bibnamefont {{Li Chen}}}, \bibinfo {author} {\bibnamefont {{Edmund H. Linfield}}}, \bibinfo {author} {\bibnamefont {{A. Giles Davies}}}, \ and\ \bibinfo {author} {\bibnamefont {{John E. Cunningham}}},\ }\bibfield  {title} {\enquote {\bibinfo {title} {{Excitation, detection and electrostatic manipulation of terahertz-frequency range plasmons in a two-dimensional electron system}},}\ }\href {\doibase 10.1038/srep15420} {\bibfield  {journal} {\bibinfo  {journal} {Sci. Rep.5}\ } (\bibinfo {year} {2015}),\ 10.1038/srep15420}\BibitemShut {NoStop}%
\bibitem [{\citenamefont {Wood}\ \emph {et~al.}(2013)\citenamefont {Wood}, \citenamefont {Mistry}, \citenamefont {Li}, \citenamefont {Cunningham}, \citenamefont {Linfield},\ and\ \citenamefont {Davies}}]{wood2013}%
  \BibitemOpen
  \bibfield  {author} {\bibinfo {author} {\bibfnamefont {C.~D.}\ \bibnamefont {Wood}}, \bibinfo {author} {\bibfnamefont {D.}~\bibnamefont {Mistry}}, \bibinfo {author} {\bibfnamefont {L.~H.}\ \bibnamefont {Li}}, \bibinfo {author} {\bibfnamefont {J.~E.}\ \bibnamefont {Cunningham}}, \bibinfo {author} {\bibfnamefont {E.~H.}\ \bibnamefont {Linfield}}, \ and\ \bibinfo {author} {\bibfnamefont {A.~G.}\ \bibnamefont {Davies}},\ }\bibfield  {title} {\enquote {\bibinfo {title} {{On-chip terahertz spectroscopic techniques for measuring mesoscopic quantum systems}},}\ }\href {\doibase 10.1063/1.4816736} {\bibfield  {journal} {\bibinfo  {journal} {Review of Scientific Instruments}\ }\textbf {\bibinfo {volume} {84}},\ \bibinfo {pages} {085101} (\bibinfo {year} {2013})}\BibitemShut {NoStop}%
\bibitem [{\citenamefont {Smith}\ \emph {et~al.}(2017)\citenamefont {Smith}, \citenamefont {Jooshesh}, \citenamefont {Zhang},\ and\ \citenamefont {Darcie}}]{Smith2017}%
  \BibitemOpen
  \bibfield  {author} {\bibinfo {author} {\bibfnamefont {R.}~\bibnamefont {Smith}}, \bibinfo {author} {\bibfnamefont {A.}~\bibnamefont {Jooshesh}}, \bibinfo {author} {\bibfnamefont {J.}~\bibnamefont {Zhang}}, \ and\ \bibinfo {author} {\bibfnamefont {T.}~\bibnamefont {Darcie}},\ }\bibfield  {title} {\enquote {\bibinfo {title} {{Photoconductive generation and detection of THz-bandwidth pulses using near-field coupling to a free-space metallic slit waveguide}},}\ }\href {\doibase 10.1364/oe.25.026492} {\bibfield  {journal} {\bibinfo  {journal} {Optics Express}\ }\textbf {\bibinfo {volume} {25}},\ \bibinfo {pages} {26492} (\bibinfo {year} {2017})}\BibitemShut {NoStop}%
\bibitem [{\citenamefont {Hunter}\ \emph {et~al.}(2015)\citenamefont {Hunter}, \citenamefont {Mayorov}, \citenamefont {Wood}, \citenamefont {Russell}, \citenamefont {Li}, \citenamefont {Linfield}, \citenamefont {Davies},\ and\ \citenamefont {Cunningham}}]{Hunter2015}%
  \BibitemOpen
  \bibfield  {author} {\bibinfo {author} {\bibfnamefont {N.}~\bibnamefont {Hunter}}, \bibinfo {author} {\bibfnamefont {A.~S.}\ \bibnamefont {Mayorov}}, \bibinfo {author} {\bibfnamefont {C.~D.}\ \bibnamefont {Wood}}, \bibinfo {author} {\bibfnamefont {C.}~\bibnamefont {Russell}}, \bibinfo {author} {\bibfnamefont {L.}~\bibnamefont {Li}}, \bibinfo {author} {\bibfnamefont {E.~H.}\ \bibnamefont {Linfield}}, \bibinfo {author} {\bibfnamefont {A.~G.}\ \bibnamefont {Davies}}, \ and\ \bibinfo {author} {\bibfnamefont {J.~E.}\ \bibnamefont {Cunningham}},\ }\bibfield  {title} {\enquote {\bibinfo {title} {{On-Chip Picosecond Pulse Detection and Generation Using Graphene Photoconductive Switches}},}\ }\href {\doibase 10.1021/nl504116w} {\bibfield  {journal} {\bibinfo  {journal} {Nano Letters}\ }\textbf {\bibinfo {volume} {15}},\ \bibinfo {pages} {1591--1596} (\bibinfo {year} {2015})}\BibitemShut {NoStop}%
\bibitem [{\citenamefont {Zhong}\ \emph {et~al.}(2008)\citenamefont {Zhong}, \citenamefont {Gabor}, \citenamefont {Sharping}, \citenamefont {Gaeta},\ and\ \citenamefont {McEuen}}]{Zhong2008}%
  \BibitemOpen
  \bibfield  {author} {\bibinfo {author} {\bibfnamefont {Z.}~\bibnamefont {Zhong}}, \bibinfo {author} {\bibfnamefont {N.~M.}\ \bibnamefont {Gabor}}, \bibinfo {author} {\bibfnamefont {J.~E.}\ \bibnamefont {Sharping}}, \bibinfo {author} {\bibfnamefont {A.~L.}\ \bibnamefont {Gaeta}}, \ and\ \bibinfo {author} {\bibfnamefont {P.~L.}\ \bibnamefont {McEuen}},\ }\bibfield  {title} {\enquote {\bibinfo {title} {{Terahertz time-domain measurement of ballistic electron resonance in a single-walled carbon nanotube}},}\ }\href {\doibase 10.1038/nnano.2008.60} {\bibfield  {journal} {\bibinfo  {journal} {Nature Nanotechnology}\ }\textbf {\bibinfo {volume} {3}},\ \bibinfo {pages} {201--205} (\bibinfo {year} {2008})}\BibitemShut {NoStop}%
\bibitem [{\citenamefont {McIver}\ \emph {et~al.}(2020)\citenamefont {McIver}, \citenamefont {Schulte}, \citenamefont {Stein}, \citenamefont {Matsuyama}, \citenamefont {Jotzu}, \citenamefont {Meier},\ and\ \citenamefont {Cavalleri}}]{McIver2020}%
  \BibitemOpen
  \bibfield  {author} {\bibinfo {author} {\bibfnamefont {J.~W.}\ \bibnamefont {McIver}}, \bibinfo {author} {\bibfnamefont {B.}~\bibnamefont {Schulte}}, \bibinfo {author} {\bibfnamefont {F.-U.}\ \bibnamefont {Stein}}, \bibinfo {author} {\bibfnamefont {T.}~\bibnamefont {Matsuyama}}, \bibinfo {author} {\bibfnamefont {G.}~\bibnamefont {Jotzu}}, \bibinfo {author} {\bibfnamefont {G.}~\bibnamefont {Meier}}, \ and\ \bibinfo {author} {\bibfnamefont {A.}~\bibnamefont {Cavalleri}},\ }\bibfield  {title} {\enquote {\bibinfo {title} {{Light-induced anomalous Hall effect in graphene}},}\ }\href {\doibase 10.1038/s41567-019-0698-y} {\bibfield  {journal} {\bibinfo  {journal} {Nature Physics}\ }\textbf {\bibinfo {volume} {16}},\ \bibinfo {pages} {38--41} (\bibinfo {year} {2020})},\ \Eprint {http://arxiv.org/abs/1811.03522} {1811.03522} \BibitemShut {NoStop}%
\bibitem [{\citenamefont {Island}\ \emph {et~al.}(2020)\citenamefont {Island}, \citenamefont {Kissin}, \citenamefont {Schalch}, \citenamefont {Cui}, \citenamefont {Haque}, \citenamefont {Potts}, \citenamefont {Taniguchi}, \citenamefont {Watanabe}, \citenamefont {Averitt},\ and\ \citenamefont {Young}}]{Island2020}%
  \BibitemOpen
  \bibfield  {author} {\bibinfo {author} {\bibfnamefont {J.~O.}\ \bibnamefont {Island}}, \bibinfo {author} {\bibfnamefont {P.}~\bibnamefont {Kissin}}, \bibinfo {author} {\bibfnamefont {J.}~\bibnamefont {Schalch}}, \bibinfo {author} {\bibfnamefont {X.}~\bibnamefont {Cui}}, \bibinfo {author} {\bibfnamefont {S.~R.~U.}\ \bibnamefont {Haque}}, \bibinfo {author} {\bibfnamefont {A.}~\bibnamefont {Potts}}, \bibinfo {author} {\bibfnamefont {T.}~\bibnamefont {Taniguchi}}, \bibinfo {author} {\bibfnamefont {K.}~\bibnamefont {Watanabe}}, \bibinfo {author} {\bibfnamefont {R.~D.}\ \bibnamefont {Averitt}}, \ and\ \bibinfo {author} {\bibfnamefont {A.~F.}\ \bibnamefont {Young}},\ }\bibfield  {title} {\enquote {\bibinfo {title} {{On-chip terahertz modulation and emission with integrated graphene junctions}},}\ }\href {\doibase 10.1063/5.0005870} {\bibfield  {journal} {\bibinfo  {journal} {Applied Physics Letters}\ }\textbf {\bibinfo {volume} {116}},\ \bibinfo {pages} {161104} (\bibinfo {year} {2020})},\ \Eprint
  {http://arxiv.org/abs/2004.02059} {2004.02059} \BibitemShut {NoStop}%
\bibitem [{\citenamefont {Wang}\ \emph {et~al.}(2023)\citenamefont {Wang}, \citenamefont {Adelinia}, \citenamefont {Chavez-Cervantes}, \citenamefont {Matsuyama}, \citenamefont {Fechner}, \citenamefont {Buzzi}, \citenamefont {Meier},\ and\ \citenamefont {Cavalleri}}]{Wang2023}%
  \BibitemOpen
  \bibfield  {author} {\bibinfo {author} {\bibfnamefont {E.}~\bibnamefont {Wang}}, \bibinfo {author} {\bibfnamefont {J.~D.}\ \bibnamefont {Adelinia}}, \bibinfo {author} {\bibfnamefont {M.}~\bibnamefont {Chavez-Cervantes}}, \bibinfo {author} {\bibfnamefont {T.}~\bibnamefont {Matsuyama}}, \bibinfo {author} {\bibfnamefont {M.}~\bibnamefont {Fechner}}, \bibinfo {author} {\bibfnamefont {M.}~\bibnamefont {Buzzi}}, \bibinfo {author} {\bibfnamefont {G.}~\bibnamefont {Meier}}, \ and\ \bibinfo {author} {\bibfnamefont {A.}~\bibnamefont {Cavalleri}},\ }\bibfield  {title} {\enquote {\bibinfo {title} {{Superconducting nonlinear transport in optically driven high-temperature K$_3$C$_{60}$}},}\ }\href {\doibase 10.1038/s41467-023-42989-7} {\bibfield  {journal} {\bibinfo  {journal} {Nature Communications}\ }\textbf {\bibinfo {volume} {14}},\ \bibinfo {pages} {7233} (\bibinfo {year} {2023})}\BibitemShut {NoStop}%
\bibitem [{\citenamefont {Adelinia}\ \emph {et~al.}(2025)\citenamefont {Adelinia}, \citenamefont {Wang}, \citenamefont {Chavez-Cervantes}, \citenamefont {Matsuyama}, \citenamefont {Fechner}, \citenamefont {Buzzi}, \citenamefont {Meier},\ and\ \citenamefont {Cavalleri}}]{Adelinia2025}%
  \BibitemOpen
  \bibfield  {author} {\bibinfo {author} {\bibfnamefont {J.~D.}\ \bibnamefont {Adelinia}}, \bibinfo {author} {\bibfnamefont {E.}~\bibnamefont {Wang}}, \bibinfo {author} {\bibfnamefont {M.}~\bibnamefont {Chavez-Cervantes}}, \bibinfo {author} {\bibfnamefont {T.}~\bibnamefont {Matsuyama}}, \bibinfo {author} {\bibfnamefont {M.}~\bibnamefont {Fechner}}, \bibinfo {author} {\bibfnamefont {M.}~\bibnamefont {Buzzi}}, \bibinfo {author} {\bibfnamefont {G.}~\bibnamefont {Meier}}, \ and\ \bibinfo {author} {\bibfnamefont {A.}~\bibnamefont {Cavalleri}},\ }\bibfield  {title} {\enquote {\bibinfo {title} {{Probing optically driven thin K$_3$C$_{60}$ films with an ultrafast voltmeter}},}\ }\href {\doibase 10.1063/4.0000295} {\bibfield  {journal} {\bibinfo  {journal} {Structural Dynamics}\ }\textbf {\bibinfo {volume} {12}},\ \bibinfo {pages} {024503} (\bibinfo {year} {2025})}\BibitemShut {NoStop}%
\bibitem [{\citenamefont {Simonovich}(2022)}]{simonovich2022}%
  \BibitemOpen
  \bibfield  {author} {\bibinfo {author} {\bibfnamefont {B.}~\bibnamefont {Simonovich}},\ }\bibfield  {title} {\enquote {\bibinfo {title} {{Coupled Transmission Lines and Crosstalk}},}\ }\href@noop {} {\bibfield  {journal} {\bibinfo  {journal} {Signal Intergrity Journal}\ } (\bibinfo {year} {2022})}\BibitemShut {NoStop}%
\bibitem [{\citenamefont {GmbH}(2019)}]{CSTmanual}%
  \BibitemOpen
  \bibfield  {author} {\bibinfo {author} {\bibfnamefont {D.~S.~D.}\ \bibnamefont {GmbH}},\ }\bibfield  {title} {\enquote {\bibinfo {title} {{CST Studio Suite}},}\ }\href@noop {} {\  (\bibinfo {year} {1998-2019})}\BibitemShut {NoStop}%
\bibitem [{\citenamefont {Kusyak}(2022)}]{Kusyak2022}%
  \BibitemOpen
  \bibfield  {author} {\bibinfo {author} {\bibfnamefont {K.}~\bibnamefont {Kusyak}},\ }\bibfield  {title} {\enquote {\bibinfo {title} {{Increasing the Bandwidth of on-Chip THz Spectroscopy}},}\ }\href@noop {} {\bibfield  {journal} {\bibinfo  {journal} {Universität Hamburg}\ } (\bibinfo {year} {2022})}\BibitemShut {NoStop}%
\bibitem [{\citenamefont {{Wheeler}}(1942)}]{Wheeler1942}%
  \BibitemOpen
  \bibfield  {author} {\bibinfo {author} {\bibfnamefont {H.~A.}\ \bibnamefont {{Wheeler}}},\ }\bibfield  {title} {\enquote {\bibinfo {title} {{Formulas for the Skin Effect}},}\ }\href {\doibase 10.1109/JRPROC.1942.232015} {\bibfield  {journal} {\bibinfo  {journal} {Procedings of the IRE}\ }\textbf {\bibinfo {volume} {30}},\ \bibinfo {pages} {412--424} (\bibinfo {year} {1942})}\BibitemShut {NoStop}%
\bibitem [{\citenamefont {Grischkowsky}\ \emph {et~al.}(1990)\citenamefont {Grischkowsky}, \citenamefont {Keiding}, \citenamefont {Exter},\ and\ \citenamefont {Fattinger}}]{Grischowsky1990}%
  \BibitemOpen
  \bibfield  {author} {\bibinfo {author} {\bibfnamefont {D.}~\bibnamefont {Grischkowsky}}, \bibinfo {author} {\bibfnamefont {S.}~\bibnamefont {Keiding}}, \bibinfo {author} {\bibfnamefont {M.~v.}\ \bibnamefont {Exter}}, \ and\ \bibinfo {author} {\bibfnamefont {C.}~\bibnamefont {Fattinger}},\ }\bibfield  {title} {\enquote {\bibinfo {title} {{Far-infrared time-domain spectroscopy with terahertz beams of dielectrics and semiconductors}},}\ }\href {\doibase 10.1364/josab.7.002006} {\bibfield  {journal} {\bibinfo  {journal} {Journal of the Optical Society of America B}\ }\textbf {\bibinfo {volume} {7}},\ \bibinfo {pages} {2006} (\bibinfo {year} {1990})}\BibitemShut {NoStop}%
\bibitem [{\citenamefont {Auston}, \citenamefont {Cheung},\ and\ \citenamefont {Smith}(1984)}]{Auston1984}%
  \BibitemOpen
  \bibfield  {author} {\bibinfo {author} {\bibfnamefont {D.~H.}\ \bibnamefont {Auston}}, \bibinfo {author} {\bibfnamefont {K.~P.}\ \bibnamefont {Cheung}}, \ and\ \bibinfo {author} {\bibfnamefont {P.~R.}\ \bibnamefont {Smith}},\ }\bibfield  {title} {\enquote {\bibinfo {title} {{Picosecond photoconducting Hertzian dipoles}},}\ }\href@noop {} {\bibfield  {journal} {\bibinfo  {journal} {Applied Physics Letters}\ } (\bibinfo {year} {1984})}\BibitemShut {NoStop}%
\bibitem [{\citenamefont {Grischkowsky}(2000)}]{Grishkowsky2000}%
  \BibitemOpen
  \bibfield  {author} {\bibinfo {author} {\bibfnamefont {D.}~\bibnamefont {Grischkowsky}},\ }\bibfield  {title} {\enquote {\bibinfo {title} {{Optoelectronic characterization of transmission lines and waveguides by terahertz time-domain spectroscopy}},}\ }\href {\doibase 10.1109/2944.902161} {\bibfield  {journal} {\bibinfo  {journal} {IEEE Journal of Selected Topics in Quantum Electronics}\ }\textbf {\bibinfo {volume} {6}},\ \bibinfo {pages} {1122--1135} (\bibinfo {year} {2000})}\BibitemShut {NoStop}%
\bibitem [{\citenamefont {Koch}\ \emph {et~al.}(2023)\citenamefont {Koch}, \citenamefont {Mittleman}, \citenamefont {Ornik},\ and\ \citenamefont {Castro-Camus}}]{koch2023}%
  \BibitemOpen
  \bibfield  {author} {\bibinfo {author} {\bibfnamefont {M.}~\bibnamefont {Koch}}, \bibinfo {author} {\bibfnamefont {D.~M.}\ \bibnamefont {Mittleman}}, \bibinfo {author} {\bibfnamefont {J.}~\bibnamefont {Ornik}}, \ and\ \bibinfo {author} {\bibfnamefont {E.}~\bibnamefont {Castro-Camus}},\ }\bibfield  {title} {\enquote {\bibinfo {title} {{Terahertz time-domain spectroscopy}},}\ }\href {\doibase 10.1038/s43586-023-00232-z} {\bibfield  {journal} {\bibinfo  {journal} {Nature Reviews Methods Primers}\ }\textbf {\bibinfo {volume} {3}},\ \bibinfo {pages} {48} (\bibinfo {year} {2023})}\BibitemShut {NoStop}%
\bibitem [{\citenamefont {Díaz}\ \emph {et~al.}(2023)\citenamefont {Díaz}, \citenamefont {Anadón}, \citenamefont {Morassi}, \citenamefont {Hehn}, \citenamefont {Lemaître},\ and\ \citenamefont {Gorchon}}]{Diaz2023}%
  \BibitemOpen
  \bibfield  {author} {\bibinfo {author} {\bibfnamefont {E.}~\bibnamefont {Díaz}}, \bibinfo {author} {\bibfnamefont {A.}~\bibnamefont {Anadón}}, \bibinfo {author} {\bibfnamefont {M.}~\bibnamefont {Morassi}}, \bibinfo {author} {\bibfnamefont {M.}~\bibnamefont {Hehn}}, \bibinfo {author} {\bibfnamefont {A.}~\bibnamefont {Lemaître}}, \ and\ \bibinfo {author} {\bibfnamefont {J.}~\bibnamefont {Gorchon}},\ }\bibfield  {title} {\enquote {\bibinfo {title} {{Calibration of terahertz sampling detectors for intense unipolar picosecond current pulses in waveguides}},}\ }\href {\doibase 10.1063/5.0169020} {\bibfield  {journal} {\bibinfo  {journal} {Applied Physics Letters}\ }\textbf {\bibinfo {volume} {123}},\ \bibinfo {pages} {141106} (\bibinfo {year} {2023})},\ \Eprint {http://arxiv.org/abs/https://pubs.aip.org/aip/apl/article-pdf/doi/10.1063/5.0169020/18150508/141106\_1\_5.0169020.pdf} {https://pubs.aip.org/aip/apl/article-pdf/doi/10.1063/5.0169020/18150508/141106\_1\_5.0169020.pdf} \BibitemShut {NoStop}%
\bibitem [{\citenamefont {Potts}(2023)}]{Pottsphd2023}%
  \BibitemOpen
  \bibfield  {author} {\bibinfo {author} {\bibfnamefont {A.~M.}\ \bibnamefont {Potts}},\ }\bibfield  {title} {\enquote {\bibinfo {title} {{On-Chip Terahertz Time Domain Spectroscopy for Sub-Diffraction van der Waals Heterostructures}},}\ }\href@noop {} {\bibfield  {journal} {\bibinfo  {journal} {University of Californica, Santa Barbara}\ } (\bibinfo {year} {2023})}\BibitemShut {NoStop}%
\end{thebibliography}%

\end{document}